\documentclass{article}

    \PassOptionsToPackage{numbers, compress}{natbib}
     \usepackage[preprint]{neurips_2022}

\usepackage[utf8]{inputenc} % allow utf-8 input
\usepackage[T1]{fontenc}    % use 8-bit T1 fonts
\usepackage{hyperref}       % hyperlinks
\usepackage{url}            % simple URL typesetting
\usepackage{booktabs}       % professional-quality tables
\usepackage{amsfonts}       % blackboard math symbols
\usepackage{nicefrac}       % compact symbols for 1/2, etc.
\usepackage{microtype}      % microtypography
\usepackage{xcolor}         % colors

\usepackage{todonotes} % todo bubbles on paper

\usepackage[draft]{minted}
\usepackage{xspace}
\usepackage{parcolumns}

\usepackage{adjustbox}
\usepackage{subcaption,graphicx}
\usepackage{lipsum}

\usepackage{wrapfig}

\usepackage{pgfplots}
\DeclareUnicodeCharacter{2212}{−}
\usepgfplotslibrary{groupplots,dateplot}
\usetikzlibrary{patterns,shapes.arrows}
\pgfplotsset{compat=newest}

\everypar{\looseness=-1} %Please try and avoid dangling words on their own lines!
\def\compactify{\itemsep=0pt \topsep=0pt \partopsep=0pt \parsep=0pt}
\let\latexusecounter=\usecounter

\newenvironment{CompactEnumerate}
  {\def\usecounter{\compactify\latexusecounter}
   \begin{enumerate}}
  {\end{enumerate}\let\usecounter=\latexusecounter}
\newenvironment{parafont}{\fontfamily{ptm}\selectfont}{}
\newcommand{\Para}[1]{\vspace{2pt}\noindent\begin{parafont}\textbf{\textit{#1}}\end{parafont}}
\newcommand{\TheMutator}{{\sc Marvolo}\xspace} 

\AtBeginEnvironment{minted}{%
  }
  
\title{Marvolo: Programmatic Data Augmentation for Practical ML-Driven Malware Detection}

\author{%
  Michael D. Wong \\
  Princeton University \\
  mikedwong@cs.princeton.edu\\
   \And
   Edward Raff \\
   Booz Allen Hamilton \\
   raff\_edward@bah.com \\
   \And
   James Holt \\
   Laboratory for Physical Sciences \\
    holt@lps.umd.edu \\
   \And
   Ravi Netravali \\
   Princeton University \\
   ravian@cs.princeton.edu \\

}

\begin{document}

\maketitle

\begin{abstract}
% Malware continues to plague large organizations and is becoming increasingly difficult to detect as malware authors are constantly looking for new ways to obfuscate and diversify their malicious code. While neural networks for malware detection have shown significant improvements over traditional signature-based detectors, security researchers and practitioners often struggle to obtain sufficiently large and comprehensive datasets for training these models. This is especially challenging for institutions like banks and governments that receive targeted malware, and thus can not collect large-scale malware. Diverse benign applications are also uniquely challenging due to copyright and licensing restrictions. 
Data augmentation has been rare in the cyber security domain due to technical difficulties in altering data in a manner that is semantically consistent with the original data. This shortfall is particularly onerous given the unique difficulty of acquiring benign and malicious training data that runs into copyright restrictions, and that institutions like banks and governments receive targeted malware that will never exist in large quantities.

We present \TheMutator, a binary mutator that programmatically grows malware (and benign) datasets in a manner that boosts the accuracy of ML-driven malware detectors. \TheMutator employs semantics-preserving code transformations that mimic the alterations that malware authors and defensive benign developers routinely make in practice
% to sidestep advances in detectors or protect considered ``trade secret'' code to apply meaningful data augmentation to the domain of malware detection. 
, allowing us to generate meaningful augmented data.
Crucially, semantics-preserving transformations also enable \TheMutator to safely propagate labels from original to newly-generated data samples without mandating expensive reverse engineering of binaries. Further, \TheMutator embeds several key optimizations that keep costs low for practitioners by maximizing the density of diverse data samples generated within a given time (or resource) budget. Experiments using wide-ranging commercial malware datasets and a recent ML-driven malware detector show that \TheMutator boosts accuracies by up to 5\%, while operating on only a small fraction (15\%) of the potential input binaries.
\end{abstract}

%%%% INTRODUCTION %%%%%

\section{Introduction}
\label{sec:intro}

% Cybersecurity is more important than ever as malware authors continue to devise not only new sophisticated attack methods, but also obfuscations to hide their malicious code. Indeed, malware now causes billions of dollars in damages every year~\cite{malware-damage-cost}. Detecting malware is notoriously difficult -- e.g., in 2020, it took 280 days on average to identify and contain a data breach~\cite{breach-report} -- and many of the affected systems are crucial to the operation of hospitals, schools, governments, universities, and other critical infrastructure~\cite{malwarebytes-report}. 

Many approaches have been developed to aid practitioners in distinguishing malicious files (i.e., malware) from benign ones. Early solutions centered on employing static or dynamic analyses of binaries to identify indicators of malicious behavior, e.g., stealing user credentials \cite{malware-in-enterprise}. However, each faces significant drawbacks: static analysis relies on manually-specified signatures that struggle to generalize to newer malware variants, while dynamic analyses bring high computational costs and virtual environments that are detectable by malicious programs (which can then fly under the radar)~\cite{ml-malware-survey}. More recently, a slew of data-driven strategies have been developed to sidestep the above issues by training ML models to distinguish between benign and malicious executables~\cite{cnn-malware, malconv, ml-malware-survey, ember, Kolter2004, tabish2009malware,Ye:2007:IIM:1281192.1281308,Fan2018,Tamersoy:2014:GAL:2623330.2623342,Bozorgi:2010:BHL:1835804.1835821,Ye:2011:CFC:2020408.2020448,Ye:2010:AMC:1835804.1835820,raff_lzjd_2017}.

Despite their promise, ML-based solutions face a significant practical challenge: obtaining representative and labeled training data is infeasible for many organizations. On the one hand, commercial datasets with these properties exist, but are unattainable for many due to financial constraints~\cite{ember,sorel}, with the licensing needed can cost \$400k/year.
On the other hand, home-grown datasets face scaling and labeling challenges, e.g., benign samples are often closed-source or copyright-protected, and labeling involves error-prone manual analysis to reverse engineer each binary. Consequently, many security practitioners and researchers only have access to small datasets that lack the heterogeneity seen in the wild~\cite{motif}. For example, we find that recent ML detectors achieve accuracies of only 60-71\% when trained on small public datasets~\cite{brazilian-malware,microsoft-malware} versus their large, commercial counterparts; such degradations are unacceptable given that single-digit accuracy improvements (and any new detected malware) are celebrated by malware analysts~\cite{evasive-malware}. We note that these small datasets are also realistic in representing the challenge applying ML to targeted or otherwise unique malware families of interst (e.g., targeted banking malware) rather than broad and indiscriminate malware.

This problem is not unique to malware detection. To overcome the challenge of small and incomprehensive datasets, many researchers have successfully resorted to data augmentation to artificially grow their datasets and has been especially prominent in the computer vision communities with techniques such as image cropping and scaling \cite{imagenet-classification,augmentation-survey,diff-augment}. In contrast, augmentation techniques for malware datasets have been almost non-existing and operate over feature representations (instead of actual data) due to the difficulty in analyzing and reasoning about raw binaries \cite{ml-malware-survey, re-process, unveiling-zeus}.

To overcome this obstacle, we present \TheMutator, a binary mutator that programmatically grows (accessible) malware datasets in a manner that directly boosts the accuracy of ML-driven malware detectors. Using \TheMutator, we show the feasibility of meaningfully applying data augmentation techniques to the domain of malware detection. The driving insight behind \TheMutator's data augmentation strategy is drawn from our analysis of binaries in high-accuracy (but difficult to access) malware datasets (\S\ref{sec:approach}). In particular, we observe that these datasets routinely contain multiple versions of a given malware file that differ based on the effects of semantics-preserving code transformations, i.e., alterations to the code that change aesthetics, but not externalized behavior~\cite{obfuscation-survey}. The reason is intuitive: producing malware requires significant effort, and once a malware binary becomes detectable, code transformations are a quick way for malware authors to preserve malicious behavior while sidestepping discernible patterns.

Building on the above observation, \TheMutator performs a wide range of semantics-preserving code transformations on existing binaries in an input dataset. Crucially, this approach naturally results in automatic (accurate) labeling of the augmented data samples. The reason is that semantics-preserving transformations inherently preserve overall code behavior. Thus, labels of pre-transformed binaries can be safely carried over to the transformed versions. To the best of our knowledge, \TheMutator is the first effort made in performing a deep-dive analysis on binaries in existing datasets to employ \textit{meaningful} data augmentation techniques to malware detection datasets.

Though conceptually straightforward, realizing \TheMutator's approach in practice is complicated by the fact that performing code transformations on binaries is time-consuming and resource-intensive. Naively decompiling, mutating, and reassembling a binary can take tens of seconds to several hours. Thus, employing this methodology on even the small datasets that practitioners have access to can take thousands of hours, with the resource expenditure foregoing any cost savings from not purchasing realistic commercial datasets. To overcome this, \TheMutator embeds two complementary optimizations that collectively maximize the utility (i.e., number of realistic and diverse data samples) of the performed transformations within a user-specified time budget: (1) \textbf{Code similarity clustering}, which clusters binaries of similar compositions and operates on only a single binary from each cluster to circumvent costly operations while preserving diverse interactions between code alterations and other sections in a binary and (2) \textbf{intermediate binary generation}, which increases the number of diverse binaries output from each pass through the pipeline using a lightweight check to determines the efficacy of outputting a binary -- based on code discrepancies from the original and previously output versions -- after each transformation.

We evaluated \TheMutator using the recent MalConv~\cite{malconv} malware detector and multiple commercially-available large/small-scale datasets, i.e., the large-scale Ember~\cite{ember} dataset, as well as a small-scale Brazilian dataset~\cite{brazilian-malware}. Overall, we find that \TheMutator boosts MalConv's accuracies by up to 6\%, with most wins coming from accurately detecting previously unseen binary families -- such scenarios are intuitively more difficult to catch, but are the primary goal of any malware detection system, highlighting the practical utility of \TheMutator. % By the nature of these families being previous unseen, they are harder to detect and the main goal of any malware detection system, meaning our approach has especially large utility to practical application.  
Further, relative to performing straightforward code transformations, \TheMutator's optimizations enable MalConv to reap these benefits while operating on 85\% fewer binaries (resulting in speedups of 79$\times$). We will open-source \TheMutator post-publication.% \rn{if acc. on unseen is high, report that}

% Ed moved ethids to checklist
% \Para{Ethical concerns.} While \TheMutator can be employed by malicious actors to obfuscate their code, they already have tools of greater functionality that run faster because malware authors have their source code. \TheMutator thus gives no new abilities to malware authors, but does enable researchers to better prepare and study such obfuscations impact on modeling and detectors --- as researchers rarely have source code, let alone a an industry scale. 
% the modified binaries are not guaranteed to be executable thus preventing \TheMutator from being used to evade existing detectors.
%%%%%% MOTIVATION %%%%%%%

\section{Background and Related Work}
\label{sec:background}

 Though prior attempts have been made in data augmentation for malware detection, they do not yet perform meaningful data augmentation. In \cite{opcode-augmentation, eda-augmentation}, data augmentation for malware detection is done by representing programs as sequences of opcodes and replacing one opcode with another without preserving semantics. Further, \cite{malware-image-augmentation} augments images generated from malware, which are known to be flawed representation \cite{malconv}. In contrast to these efforts, \TheMutator's contributions lie in (1) a deep-dive analysis of large-scale malware datasets to uncover the usage patterns of semantics-preserving code transformations by malware authors, and (2) a system that leverages those insights to efficiently grow small datasets into larger ones with improved heterogeneity and realism that aid end-to-end ML-based malware detection.

% \subsection{Existing Malware Detectors}

% \Para{Primer on PE format.} Most approaches in malware detection analyze raw executables (or binaries)~\cite{malconv, ghidra, yara, cuckoo-detection}. Due to the widespread usage of Windows systems and the large amounts of malware targeting these systems, these malware detectors focus primarily on analyzing PE32 executables. As a brief primer, a PE file consists of a data structure that provides the OS with the necessary information to load the program into memory. It contains a series of headers and sections with different data. Most notably, the \texttt{.text} section contains the instructions to be executed, the \texttt{.data} section contains global variables, and the \texttt{.rsrc} section contains resources used by the program such as icons. We refer the interested reader to the PE specification for further details~\cite{pe-overview}.

Due to its prevalence, and greater difficulty, our work focuses on Microsoft Windows Portable Executable (PE) malware. Decades of development and compilation to machine code make PE processing and parsing highly non-trivial~\cite{pe-overview}, resulting in many levels of processing done by malware detectors to trade off speed and accuracy~\cite{malconv, ghidra, yara, cuckoo-detection}.

% \Para{Programmatic malware detection.}
Malware detection involves both static and dynamic analysis techniques~\cite{ml-malware-survey}. Classic Static analysis approaches primarily involved using a tool such as Yara~\cite{yara} to generate specific rules or patterns for identifying malicious files. However, static signatures fail to keep pace with the rapidly evolving space of deployed malware variants \cite{evasive-malware} and can take days of manual effort~\cite{autoyara,Votipka2019}. Malware detectors rooted in dynamic analysis~\cite{dynamic-survey} execute a binary in a sandbox to observe its behavior while restricting potential damage. Dynamic approaches step past the limitations of static analyses, the required analysis can be computationally expensive because each file often must be executed multiple times to elicit harmful behavior. Worse, some malicious binaries embed checks to detect whether they are running a virtual (sandbox) environment based on VM properties such as the amount of available DRAM, the number of cores, the list of installed applications/tools, and even the temperature of the CPU~\cite{ml-malware-survey} and dynamically alter their behavior to evade detection.
% We note that our approach is the only method that can be used for data augmentation across this spectrum of detector types, but our testing will be restrained to one static model to prove effectiveness and due to compute limitations. 

% \Para{Data-driven malware detection.} 
To address the above limitations and deliver detection accuracy (and generalization), data-driven techniques using deep learning models have seen significant traction in recent years. These models typically consist of neural networks that determine whether or not a given binary is malicious or benign based on various, defining features of that binary. For instance, certain models run inference over PE header values, assembly code, network traffic, and even the names of binaries~\cite{ml-malware-survey,names-malware}. Others follow a dynamic approach and perform manual feature engineering of API calls~\cite{dynamic-ml-model}. Most recently, the MalConv CNN~\cite{malconv} performs malware detection by operating directly over the raw bytes in a binary, thereby eschewing labor-intensive feature engineering and the need for domain expertise.

\subsection{The Problem: Limited (Realistic) Data}
\label{sec:background:the-problem}

The effectiveness of data-driven malware detectors heavily depends on the data used to train the corresponding neural networks. Unfortunately, to date, it is practically difficult for practitioners to obtain access to training datasets that are sufficiently representative of malware in the wild.

Commercial datasets that contain massive amounts of labeled data samples for malware detection do exist and have been used to train models that deliver excellent malware detection accuracy in the wild~\cite{ember}. For instance, the popular Ember dataset contains 1.1 million samples and close to 3,000 distinct malware families. However, obtaining the raw executables in the Ember dataset mandates having a VirusTotal license, which can cost upwards of \$400,000 per year!\footnote{Ember's free offering omits executables, and only presents a limited number of features per binary, e.g., size, library functions. These features are insufficient for most existing data-driven malware detectors, and cannot support long term development: analysts must avoid having adversaries learn about the used features, and cannot test new features without access to the binaries.} While there are datasets consisting of raw malware binaries \cite{sorel}, they do not contain benign binaries. Thus, practitioners must download only a limited number of binaries to prevent data imbalance which is far from trivial since it is difficult to determine a priori which binaries will allow the model to generalize. Consequently, many cost-constrained practitioners and research groups must resort to far smaller datasets that are publicly available, e.g., the Brazilian malware dataset contains 50K files~\cite{brazilian-malware}, while the Microsoft malware dataset contains 20K files with only 9 malware families~\cite{microsoft-malware}.

On the other hand, practitioners can opt to generate homegrown datasets using honeypots that attract malware binaries~\cite{intrusion-detection-honeypots}. However, such approaches face three challenges. First, the type of malware that is gathered is dependent on the collection methodology set by the user, leading to biased datasets~\cite{ml-malware-survey}. Second, collecting a sufficient number of benign data samples is difficult as benignware does not seek to replicate across machines (like malware does), and software is often closed-source and copyright-protected. Along these lines, many seminal works in malware detection have struggled to obtain benign executables, often collecting them from clean installations \cite{Kolter2004,tabish2009malware}, but this fails to obtain more than a few thousand samples. More recent works often rely on partnerships with anti-virus companies in order to obtain sufficient benign samples \cite{Fan2018,Tamersoy:2014:GAL:2623330.2623342,Bozorgi:2010:BHL:1835804.1835821,Ye:2011:CFC:2020408.2020448,Ye:2010:AMC:1835804.1835820,raff_lzjd_2017}. This naturally results in unsharable data, causing reproducibility challenges~\cite{ml-malware-survey}, slows research by non-connected groups, and neglects the needs of niche and targeted malware\cite{autoyara,brazilian-malware}. Finally, even if practitioners were to obtain a large number of samples, labeling them is not straightforward. Software reverse-engineering tools exist~\cite{ghidra}, but can consume many hours to reverse engineer a single executable, even for expert analysts~\cite{ml-malware-survey, re-process, unveiling-zeus}.

% Full refs: \cite{Fan2018,Ding:2016:KMA:2939672.2939719,Ye:2007:IIM:1281192.1281308,Tamersoy:2014:GAL:2623330.2623342,Bozorgi:2010:BHL:1835804.1835821,10.1145/2939672.2939765,Ye:2011:CFC:2020408.2020448,Ye:2010:AMC:1835804.1835820,10.1145/3447548.3467168,raff_lzjd_2017}

% \begin{figure}[!t]
\begin{wrapfigure}[11]{r}{0.41\textwidth}
\vspace{-10pt}
\includegraphics[width=0.4\columnwidth,height=0.22\columnwidth]{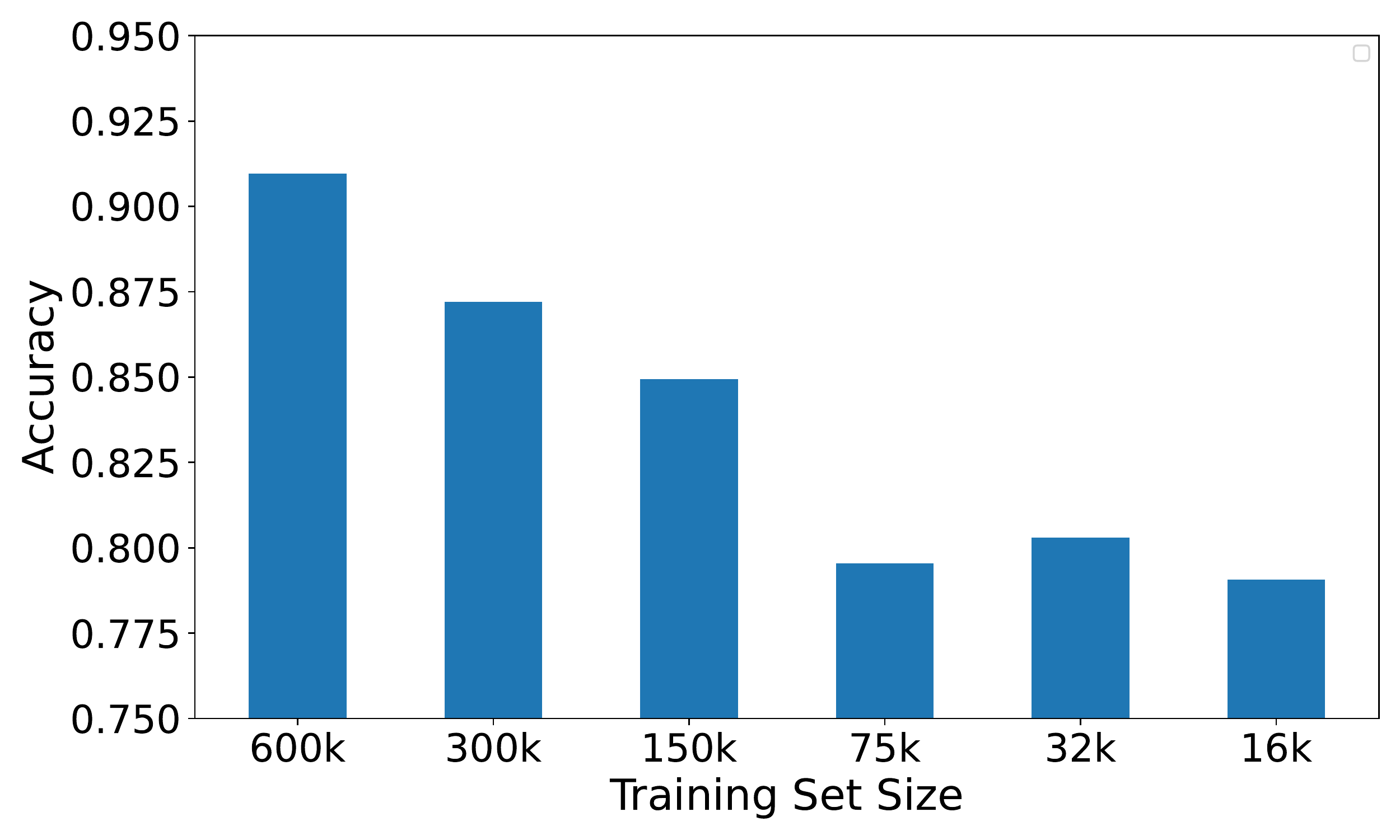}
\caption{Accuracy when training MalConv~\cite{malconv} on different subsets of the Ember dataset~\cite{ember}.}
\label{fig:small-dataset}
% \end{figure}
\end{wrapfigure}

To demonstrate the sensitivity of ML-based malware detectors to dataset composition and size, we ran experiments comparing the efficacy of models trained with commercial large-scale (Ember) and small-scale datasets (the public Microsoft and Brazilian datasets described above). Results use the recent MalConv detector~\cite{malconv}, and follow the setup described in \S\ref{sec:evaluation} (testing is done on the 200K Ember test set). To contextualize these results, we note that the implications of detecting even a single additional malicious binary in the wild can be substantial (\S\ref{sec:intro}), and that single-digit accuracy improvements are celebrated by malware analysts~\cite{evasive-malware}.
% TODO: Better citation for this?

\Para{Takeaway 1: small malware datasets lack heterogeneity, fail to generalize.} Across the considered free, small datasets that are sized between 20-75k samples, MalConv's accuracy spanned only 60-71\% relative to a training on the full Ember training dataset (600k).

\Para{Takeaway 2: large (proven) malware datasets have important diversity that detectors capitalize on.} Figure~\ref{fig:small-dataset} shows the diminishing accuracy of MalConv when trained on progressively fewer data samples from the Ember dataset. Starting with the full 600K Ember training dataset, accuracy is at 91\%. However, accuracy dips below 80\% when trained on subsets sized similarly to existing free datasets, e.g., 75k samples and less. These results indicate the data-hungry nature of ML-based malware detectors, and highlight the heterogeneity in data samples in large datasets; we dig deeper into these aspects in the following section.

%%%% APPROACH %%%%%
\section{Approach}
 \label{sec:approach}
 
Our results from Section~\ref{sec:background} highlight the inadequacies of small malware datasets relative to the large (commercial) datasets that have supported high accuracies for ML-driven malware detectors in practical settings. However, given the superior attainability of small datasets, our main goal is to determine whether they can be altered to more closely mimic the properties of their larger counterparts and deliver similar efficacy when used to train malware detectors. To do so, we programmatically analyzed the binaries in the large Ember dataset to identify their defining characteristics. We start with representative case studies that illustrate our findings, before describing more general takeaways.

% \begin{figure}[t]
\begin{wrapfigure}[11]{r}{0.54\textwidth}
\vspace{-10pt}
% \hspace{-20pt}
    \begin{minipage}{0.10\columnwidth}
        \centering
        \underline{\textbf{Zenpak}}
\begin{minted}[fontsize=\scriptsize]{gas}
inc eax
inc ecx
inc edx
inc ebx
inc esp
inc ebp
inc esi
inc edi
\end{minted}
\end{minipage}
\begin{minipage}{0.10\columnwidth}
\centering
\phantom{Zenpack}
\begin{minted}[fontsize=\scriptsize]{gas}
dec eax
dec ecx
dec edx
dec ebx
dec esp
dec ebp
dec esi
dec edi
\end{minted}
    \end{minipage}
    \hspace{-12pt}
    \begin{minipage}{0.13\columnwidth}
        \centering
        \underline{\textbf{Sivis}}
         \begin{minted}[fontsize=\scriptsize]{gas}
    nop
    nop
    nop
    xor eax, eax
    inc ebx
    dec ebx
    inc ecx
    dec ecx
\end{minted}
\end{minipage}
\begin{minipage}{0.20\columnwidth}
\centering
\phantom{Sivis}
\begin{minted}[fontsize=\scriptsize]{gas}
    inc eax
    push edx
    xor edx, edx
    pop edx
    inc eax
    dec eax
    cmp 0x17b8ef93, eax
    jne 0x407033
        \end{minted}
    \end{minipage}
    \caption{Code snippets from two malware families in the Ember dataset that exhibit semantics-preserving code transformations.}
    \label{fig:obfuscated-code}
% \end{figure}
\end{wrapfigure}

\Para{Case study I.} Figure \ref{fig:obfuscated-code} shows code snippets from two different malware families in the Ember dataset: the Zenpak malware family, and the Sivis malware family.\footnote{x86 assembly code samples are written in Intel syntax.} The first binary from Zenpak uses a code obfuscation technique called junk code insertion~\cite{obfuscation-survey}. Junk code is comprised of instructions that are executed but do not affect the externalized output(s) of the program. Here, junk code manifests as a series of \texttt{inc} instructions (line 1-8) that each increment a register's value, immediately followed by \texttt{dec} instructions (lines 9-16) that decrement them.

The binary from Sivis also uses multiple forms of junk code insertion: (1) the \texttt{nop} instructions (lines 1-3) which do not trigger any computation or data movement, (2) the interleaved \texttt{inc} and \texttt{dec} that sequentially alter the same registers (lines 5-8, 13-14), and (3) lines 10-12 which \texttt{push} the value of \texttt{edx} onto the stack, set the value of \texttt{edx} to 0 using \texttt{xor}, and then pop the old value of \texttt{edx} from the stack and store it back into \texttt{edx} (rendering the \texttt{xor} operation useless). The Sivis binary embeds another code obfuscation technique called opaque predicates~\cite{obfuscation-survey}, which are (typically) known a priori by a programmer to always evaluate to true or false. This manifests in relation to \texttt{eax}. At the start of the snippet, \texttt{eax} is definitively set to 0 after the \texttt{xor} instruction (line 4). However, at the point of the \texttt{cmp} instruction in line 15, the value stored in \texttt{eax} is definitively 1 due to the series of \texttt{inc} and \texttt{dec} operations in the preceding statements. In line 15, since \texttt{eax} $\neq$ \texttt{0x17b8ef93}, the jump in the following \texttt{jne} instruction is always taken.

\begin{wrapfigure}[12]{r}{0.56\textwidth}
\vspace{-10pt}
% \begin{figure}[!h]
    \begin{minipage}{0.3\columnwidth}
        %\centering
        \underline{\textbf{Binary 1}}
        \begin{minted}[breaklines,fontsize=\scriptsize]{gas}
push ebx
push esi
mov esi,DWORD PTR [ebp+0x8]
push edi
mov eax,ds:0x470208
push 0x7 
pop ecx
lea edi,DWORD PTR [ebp-0x2c]
    \end{minted}
    \end{minipage}
    % \hspace{40pt}
    \begin{minipage}[breaklines]{0.25\columnwidth}
        %\centering
        \underline{\textbf{Binary 2}}
         \begin{minted}[fontsize=\scriptsize]{gas}
mov eax,ds:0x423e88
push ebx
push esi
mov esi,DWORD PTR [ebp+0x8]
push edi
push 0x7 
pop ecx
lea edi,DWORD PTR [ebp-0x28]
    \end{minted}
    \end{minipage}
    \caption{Snippets from two binaries in the same ``InstallMonster'' family that exhibit minor differences due to code obfuscations.}
    \label{fig:installmonster}
% \end{figure}
\end{wrapfigure}
\Para{Case study II.} Figure~\ref{fig:installmonster} depicts snippets from two sample binaries from the Ember dataset that belong to the same family. Unsurprisingly, the two code snippets are similar at first glance. However, there exist minor differences due to two code obfuscation techniques that they embed. First, each binary uses a \texttt{mov} instruction to write data from the data segment into \texttt{eax}. However, the data is located in different memory locations across the two version; the two binaries retrieve the value from \texttt{ds:0x470208} and \texttt{ds:0x324e88}, respectively. This pattern is also seen in the \texttt{lea} instructions where the two binaries use different offsets from the stack base pointer, \texttt{ebp}, to retrieve their values. In addition, the two binaries use instruction swapping to reorder instructions (in this case, the \texttt{mov} instruction) in a manner that preserves overall semantics.

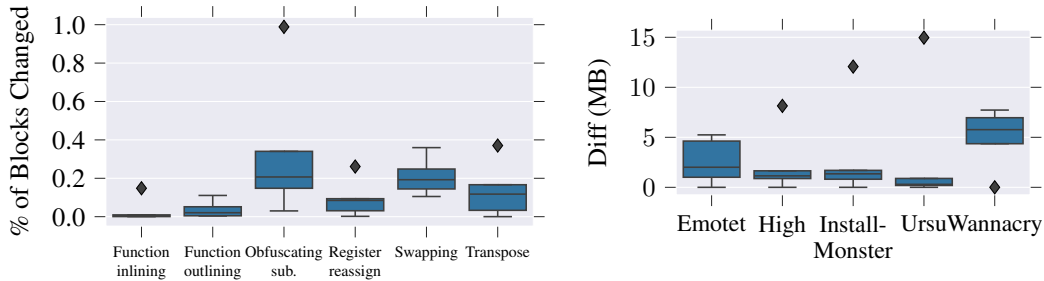
\begin{figure}[!h]
  \centering
  \begin{subfigure}{0.52\linewidth}
    % This file was created with tikzplotlib v0.9.14.
\begin{tikzpicture}

\definecolor{color0}{rgb}{0.917647058823529,0.917647058823529,0.949019607843137}
\definecolor{color1}{rgb}{0.194607843137255,0.453431372549019,0.632843137254902}

\begin{axis}[
axis background/.style={fill=color0},
axis line style={white},
tick align=outside,
tick pos=both,
x grid style={white},
xmin=-0.5, xmax=5.5,
xtick style={color=white!15!black},
xtick={0,1,2,3,4,5},
xticklabel style = {font=\tiny,align=center},
xticklabels={
  Function\\inlining,
  Function\\outlining,
  Obfuscating\\sub.,
  Register\\reassign,
  Swapping,
  Transpose
},
y grid style={white},
ylabel={\% of Blocks Changed},
ymajorgrids,
ymin=-0.0492159575, ymax=1.0379670075,
ytick style={color=white!15!black},
ytick={-0.2,0,0.2,0.4,0.6,0.8,1,1.2},
yticklabels={−0.2,0.0,0.2,0.4,0.6,0.8,1.0,1.2},
width=\columnwidth,
height=0.6\columnwidth
]
\path [draw=white!24.7058823529412!black, fill=color1, semithick]
(axis cs:-0.4,0.0005984)
--(axis cs:0.4,0.0005984)
--(axis cs:0.4,0.009324)
--(axis cs:-0.4,0.009324)
--(axis cs:-0.4,0.0005984)
--cycle;
\path [draw=white!24.7058823529412!black, fill=color1, semithick]
(axis cs:0.6,0.00465)
--(axis cs:1.4,0.00465)
--(axis cs:1.4,0.05162)
--(axis cs:0.6,0.05162)
--(axis cs:0.6,0.00465)
--cycle;
\path [draw=white!24.7058823529412!black, fill=color1, semithick]
(axis cs:1.6,0.1479977)
--(axis cs:2.4,0.1479977)
--(axis cs:2.4,0.3403242)
--(axis cs:1.6,0.3403242)
--(axis cs:1.6,0.1479977)
--cycle;
\path [draw=white!24.7058823529412!black, fill=color1, semithick]
(axis cs:2.6,0.030936)
--(axis cs:3.4,0.030936)
--(axis cs:3.4,0.093608)
--(axis cs:2.6,0.093608)
--(axis cs:2.6,0.030936)
--cycle;
\path [draw=white!24.7058823529412!black, fill=color1, semithick]
(axis cs:3.6,0.1444652)
--(axis cs:4.4,0.1444652)
--(axis cs:4.4,0.248151)
--(axis cs:3.6,0.248151)
--(axis cs:3.6,0.1444652)
--cycle;
\path [draw=white!24.7058823529412!black, fill=color1, semithick]
(axis cs:4.6,0.0331491)
--(axis cs:5.4,0.0331491)
--(axis cs:5.4,0.16642599)
--(axis cs:4.6,0.16642599)
--(axis cs:4.6,0.0331491)
--cycle;
\addplot [semithick, white!24.7058823529412!black]
table {%
0 0.0005984
0 0.00020584602
};
\addplot [semithick, white!24.7058823529412!black]
table {%
0 0.009324
0 0.009324
};
\addplot [semithick, white!24.7058823529412!black]
table {%
-0.2 0.00020584602
0.2 0.00020584602
};
\addplot [semithick, white!24.7058823529412!black]
table {%
-0.2 0.009324
0.2 0.009324
};
\addplot [black, mark=diamond*, mark size=2.5, mark options={solid,fill=white!24.7058823529412!black}, only marks]
table {%
0 0.147997
};
\addplot [semithick, white!24.7058823529412!black]
table {%
1 0.00465
1 0.0037243
};
\addplot [semithick, white!24.7058823529412!black]
table {%
1 0.05162
1 0.11061
};
\addplot [semithick, white!24.7058823529412!black]
table {%
0.8 0.0037243
1.2 0.0037243
};
\addplot [semithick, white!24.7058823529412!black]
table {%
0.8 0.11061
1.2 0.11061
};
\addplot [semithick, white!24.7058823529412!black]
table {%
2 0.1479977
2 0.030075
};
\addplot [semithick, white!24.7058823529412!black]
table {%
2 0.3403242
2 0.3403242
};
\addplot [semithick, white!24.7058823529412!black]
table {%
1.8 0.030075
2.2 0.030075
};
\addplot [semithick, white!24.7058823529412!black]
table {%
1.8 0.3403242
2.2 0.3403242
};
\addplot [black, mark=diamond*, mark size=2.5, mark options={solid,fill=white!24.7058823529412!black}, only marks]
table {%
2 0.9885496
};
\addplot [semithick, white!24.7058823529412!black]
table {%
3 0.030936
3 0.001844
};
\addplot [semithick, white!24.7058823529412!black]
table {%
3 0.093608
3 0.093608
};
\addplot [semithick, white!24.7058823529412!black]
table {%
2.8 0.001844
3.2 0.001844
};
\addplot [semithick, white!24.7058823529412!black]
table {%
2.8 0.093608
3.2 0.093608
};
\addplot [black, mark=diamond*, mark size=2.5, mark options={solid,fill=white!24.7058823529412!black}, only marks]
table {%
3 0.2611444
};
\addplot [semithick, white!24.7058823529412!black]
table {%
4 0.1444652
4 0.1051513
};
\addplot [semithick, white!24.7058823529412!black]
table {%
4 0.248151
4 0.359678
};
\addplot [semithick, white!24.7058823529412!black]
table {%
3.8 0.1051513
4.2 0.1051513
};
\addplot [semithick, white!24.7058823529412!black]
table {%
3.8 0.359678
4.2 0.359678
};
\addplot [semithick, white!24.7058823529412!black]
table {%
5 0.0331491
5 0.00020145
};
\addplot [semithick, white!24.7058823529412!black]
table {%
5 0.16642599
5 0.16642599
};
\addplot [semithick, white!24.7058823529412!black]
table {%
4.8 0.00020145
5.2 0.00020145
};
\addplot [semithick, white!24.7058823529412!black]
table {%
4.8 0.16642599
5.2 0.16642599
};
\addplot [black, mark=diamond*, mark size=2.5, mark options={solid,fill=white!24.7058823529412!black}, only marks]
table {%
5 0.3705207
};
\addplot [semithick, white!24.7058823529412!black]
table {%
-0.4 0.003590757
0.4 0.003590757
};
\addplot [semithick, white!24.7058823529412!black]
table {%
0.6 0.0206751
1.4 0.0206751
};
\addplot [semithick, white!24.7058823529412!black]
table {%
1.6 0.20683287165
2.4 0.20683287165
};
\addplot [semithick, white!24.7058823529412!black]
table {%
2.6 0.084709
3.4 0.084709
};
\addplot [semithick, white!24.7058823529412!black]
table {%
3.6 0.1925805
4.4 0.1925805
};
\addplot [semithick, white!24.7058823529412!black]
table {%
4.6 0.1175222
5.4 0.1175222
};
\end{axis}

\end{tikzpicture} 
    \caption{Percentages of code blocks in Ember's binaries that are affected by different code transformations.}
    \label{fig:blocks-affected}
  \end{subfigure}
  \hfill
    \begin{subfigure}{0.45\linewidth}
% This file was created with tikzplotlib v0.9.14.
\begin{tikzpicture}

\definecolor{color0}{rgb}{0.917647058823529,0.917647058823529,0.949019607843137}
\definecolor{color1}{rgb}{0.194607843137255,0.453431372549019,0.632843137254902}

\begin{axis}[
axis background/.style={fill=color0},
axis line style={white},
tick align=outside,
tick pos=both,
x grid style={white},
xmin=-0.5, xmax=4.5,
xtick style={color=white!15!black},
xtick={0,1,2,3,4},
xticklabel style = {font=\small,align=center},
xticklabels={Emotet,High,Install-\\Monster,Ursu,Wannacry},
y grid style={white},
ylabel={Diff (MB)},
ymajorgrids,
ymin=-0.74819405, ymax=15.71211905,
ytick style={color=white!15!black},
width=\columnwidth,
height=0.6\columnwidth
]
\path [draw=white!24.7058823529412!black, fill=color1, semithick]
(axis cs:-0.4,1.007004)
--(axis cs:0.4,1.007004)
--(axis cs:0.4,4.622404)
--(axis cs:-0.4,4.622404)
--(axis cs:-0.4,1.007004)
--cycle;
\path [draw=white!24.7058823529412!black, fill=color1, semithick]
(axis cs:0.6,0.87637)
--(axis cs:1.4,0.87637)
--(axis cs:1.4,1.649488)
--(axis cs:0.6,1.649488)
--(axis cs:0.6,0.87637)
--cycle;
\path [draw=white!24.7058823529412!black, fill=color1, semithick]
(axis cs:1.6,0.809019)
--(axis cs:2.4,0.809019)
--(axis cs:2.4,1.694686)
--(axis cs:1.6,1.694686)
--(axis cs:1.6,0.809019)
--cycle;
\path [draw=white!24.7058823529412!black, fill=color1, semithick]
(axis cs:2.6,0.188477)
--(axis cs:3.4,0.188477)
--(axis cs:3.4,0.886716)
--(axis cs:2.6,0.886716)
--(axis cs:2.6,0.188477)
--cycle;
\path [draw=white!24.7058823529412!black, fill=color1, semithick]
(axis cs:3.6,4.359819)
--(axis cs:4.4,4.359819)
--(axis cs:4.4,6.958665)
--(axis cs:3.6,6.958665)
--(axis cs:3.6,4.359819)
--cycle;
\addplot [semithick, white!24.7058823529412!black]
table {%
0 1.007004
0 7e-06
};
\addplot [semithick, white!24.7058823529412!black]
table {%
0 4.622404
0 5.243961
};
\addplot [semithick, white!24.7058823529412!black]
table {%
-0.2 7e-06
0.2 7e-06
};
\addplot [semithick, white!24.7058823529412!black]
table {%
-0.2 5.243961
0.2 5.243961
};
\addplot [semithick, white!24.7058823529412!black]
table {%
1 0.87637
1 2e-06
};
\addplot [semithick, white!24.7058823529412!black]
table {%
1 1.649488
1 1.649488
};
\addplot [semithick, white!24.7058823529412!black]
table {%
0.8 2e-06
1.2 2e-06
};
\addplot [semithick, white!24.7058823529412!black]
table {%
0.8 1.649488
1.2 1.649488
};
\addplot [black, mark=diamond*, mark size=2.5, mark options={solid,fill=white!24.7058823529412!black}, only marks]
table {%
1 8.141279
};
\addplot [semithick, white!24.7058823529412!black]
table {%
2 0.809019
2 0.000254
};
\addplot [semithick, white!24.7058823529412!black]
table {%
2 1.694686
2 1.694686
};
\addplot [semithick, white!24.7058823529412!black]
table {%
1.8 0.000254
2.2 0.000254
};
\addplot [semithick, white!24.7058823529412!black]
table {%
1.8 1.694686
2.2 1.694686
};
\addplot [black, mark=diamond*, mark size=2.5, mark options={solid,fill=white!24.7058823529412!black}, only marks]
table {%
2 12.069082
};
\addplot [semithick, white!24.7058823529412!black]
table {%
3 0.188477
3 1e-05
};
\addplot [semithick, white!24.7058823529412!black]
table {%
3 0.886716
3 0.886716
};
\addplot [semithick, white!24.7058823529412!black]
table {%
2.8 1e-05
3.2 1e-05
};
\addplot [semithick, white!24.7058823529412!black]
table {%
2.8 0.886716
3.2 0.886716
};
\addplot [black, mark=diamond*, mark size=2.5, mark options={solid,fill=white!24.7058823529412!black}, only marks]
table {%
3 14.963923
};
\addplot [semithick, white!24.7058823529412!black]
table {%
4 4.359819
4 4.359819
};
\addplot [semithick, white!24.7058823529412!black]
table {%
4 6.958665
4 7.723609
};
\addplot [semithick, white!24.7058823529412!black]
table {%
3.8 4.359819
4.2 4.359819
};
\addplot [semithick, white!24.7058823529412!black]
table {%
3.8 7.723609
4.2 7.723609
};
\addplot [black, mark=diamond*, mark size=2.5, mark options={solid,fill=white!24.7058823529412!black}, only marks]
table {%
4 0.001092
};
\addplot [semithick, white!24.7058823529412!black]
table {%
-0.4 2.00758
0.4 2.00758
};
\addplot [semithick, white!24.7058823529412!black]
table {%
0.6 1.146014
1.4 1.146014
};
\addplot [semithick, white!24.7058823529412!black]
table {%
1.6 1.348231
2.4 1.348231
};
\addplot [semithick, white!24.7058823529412!black]
table {%
2.6 0.298814
3.4 0.298814
};
\addplot [semithick, white!24.7058823529412!black]
table {%
3.6 5.766237
4.4 5.766237
};
\end{axis}

\end{tikzpicture}
\caption{Pairwise byte diff results between binaries in five representative malware families.}
\label{fig:pairwise_diffs}
  \end{subfigure}
  \caption{Left shows percentage of code blocks that are mutated by each transformation type. Right the difference in file size for a subset of representative mlaware families.}\label{fig:pairwise_diffs} \label{fig:blocks-affected}
 \hfill
\end{figure}

% \Para{Takeaways.} 
Our case studies highlight two main points (which we repeatedly observed across the Ember dataset):
% \begin{CompactEnumerate}

\indent (1) \Para{Semantics-preserving code transformations.} Malware authors routinely alter prior versions of malicious programs using code obfuscation techniques that preserve program behavior. The reason is intuitive: generating malware involves much manual labor and sophisticated code alteration. As malware detectors discern already-deployed malware by recognizing patterns in their code composition or execution regimes (\S\ref{sec:background}), a far less challenging way for malware authors to continue deploying their malicious code is to perform semantics-preserving code transformations. More specifically, these transformations alter that code minimally, so as to preserve its malicious behavior while deviating from the patterns used to detect its predecessor. Unsurprisingly, we did not observe any remnants of semantics-preserving code transformations in the benign samples that we analyzed. 

%we did not detect any obfuscations, which is not surprising since obfuscations are more commonly used by malware authors to bypass anti-virus detectors and make reverse-engineering by analysts difficult. 

\indent (2) \Para{Combinations of transformations.} To ensure sufficient differences from detected malware versions, malware authors often resort to performing semantics-preserving transformations, e.g., as in case study II above. This approach is fruitful as such transformations are often (logically) complementary, and the effect of each transformation depends on subtle interactions between the transformation logic and binary code (ranges shown in Figure~\ref{fig:blocks-affected}). Additionally, we find that, to further boost diversity with multiple transformations, each obfuscation is not necessarily applied to all possible blocks in a binary, i.e., some binaries exhibited the effects of an obfuscation in all code blocks that it applied to, while others demonstrated the effects in only a fraction of those blocks.

Taking a step back, these observations lead to two implications about the large datasets that have been successfully used for ML-driven malware detection. First, there exist far fewer families of malicious binaries than malicious binaries themselves; the Ember dataset includes 300K malicious binary samples spread across only 332 families. 
% Accordingly, as shown in Figure~\ref{fig:family-size-cdf}, there 
There exist many binary versions per family: there are 287 and 13,951 binaries in the median and 99th percentile families, respectively. Second, the binaries within each family can differ quite substantially depending on the specific transformations that are applied across versions. Figure~\ref{fig:pairwise_diffs} highlights this property, showing that for subsets of five representative families, the constituent binaries exhibit median pairwise percent differences of 38-99\% (which equates to raw differences of 0.8--5.4 MB).

\Para{Our approach.} The results above motivate a new approach to bolstering the efficacy of the small datasets that practitioners are often restricted to: data augmentation via semantics-preserving transformations. In other words, we aim to grow small datasets by performing different combinations of semantics-preserving code transformations on varying numbers of blocks in the constituent binaries. Doing so mimics the techniques that malware authors use to sidestep malware detectors over time \cite{evasive-malware}, and yield data similar to that in (proven) large datasets. We employ further code transformations done by optimizing compilers to generate new benign binaries. Perhaps more importantly, semantics-preserving transformations provide a direct path to accurately labeling newly generated data without manual effort since pre- and post-transformation binaries will exhibit the same behavior (and thus can safely share labels). \S\ref{sec:marvolo} describes how our system, \TheMutator, practically realizes this approach.

\begin{figure}[!h]
  \centering
  \begin{subfigure}{0.5\linewidth}
    \includegraphics[width=0.8\columnwidth]{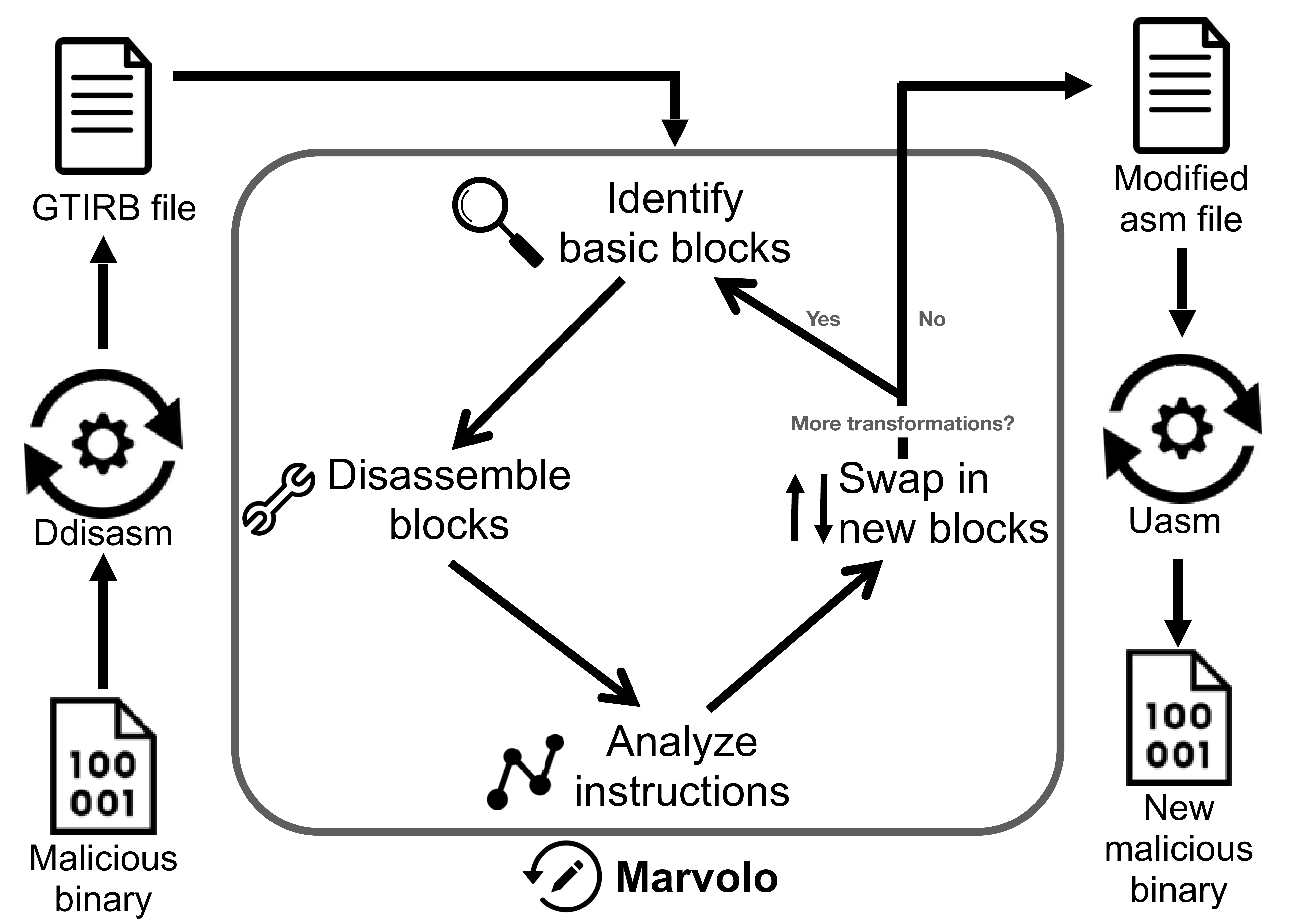}
    % \caption{\TheMutator workflow for mutating a malicious binary.}
    \label{fig:workflow}
  \end{subfigure}
  \hfill
    \begin{subfigure}{0.45\linewidth}
    \includegraphics[width=0.8\columnwidth]{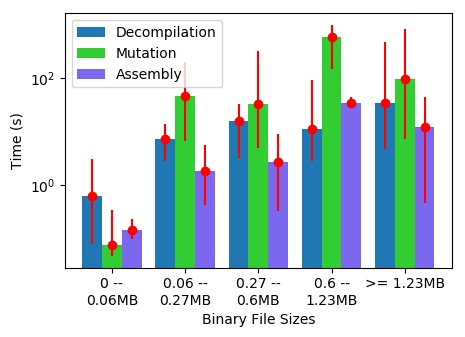}
    % \caption{Breakdown of time spent on each stage in \TheMutator's pipeline (Figure~\ref{fig:workflow}) for (a single run on) binaries in different size groups. Bars list medians, with error bars for 25-75th percentiles.}
    \label{fig:total-pipeline}
  \end{subfigure}
  \caption{\TheMutator workflow shown in (left), and the breakdown of time spent on each state of the pipeline in (right) for binaries of different sizes. Bars list medians with errors bars for the quartiles.}\label{fig:workflow}
    \label{fig:total-pipeline}
 \hfill
\end{figure}

%%%%%% MARVOLO %%%%%%
\section{\TheMutator}
\label{sec:marvolo}

% In this section, we detail \TheMutator's workflow, including the binary rewriting process, the code transformations that \TheMutator performs, and optimizations to enhance practicality. 
%and briefly discuss the correctness of these transformations.

\subsection{Binary rewriting overview}

% Figure \ref{fig:workflow} illustrates \TheMutator's binary mutation process for performing semantics-preserving transformations on a single (malicious) binary. To begin mutation, \TheMutator decompiles existing PE32 binaries using the Ddisasm tool \cite{ddisasm}, which outputs a GTIRB file~\cite{gtirb}. GTIRB is an intermediate representation that captures the general, high-level structure of the binary and stores the raw instruction bytes (but not the semantics of those instructions). \TheMutator ingests the GTIRB file and internally represents the binary as a series of basic instruction (or code) \emph{blocks}.
Figure \ref{fig:workflow} illustrates \TheMutator's binary mutation process for performing semantics-preserving transformations on a single (malicious) binary. To begin mutation, \TheMutator decompiles existing PE32 binaries using the Ddisasm tool \cite{ddisasm} and internally represents the binary as a series of basic instruction (or code) \emph{blocks}. 

To operate on (i.e., mutate) instruction blocks, \TheMutator first disassembles each block. The resulting blocks are then passed into the \TheMutator code transformation engine, which (1) selects a set of semantics-preserving code transformations to apply to the binary during a given iteration, (2) analyzes all blocks to determine which blocks each considered transformation is applicable to, (3)  selects the fraction of potential blocks to apply each transformation to, and (4) sequentially carries out the transformations on the selected blocks; \S\ref{sec:code_transformations} details this process. After code transformations are complete for a given iteration, \TheMutator then directly swaps out the corresponding (unmodified) blocks with their transformed counterparts and invokes an assembler to get the output binary. This binary is then added to the original dataset and tagged with the same label (i.e., malicious or benign) as the one used during its generation. This end-to-end process repeats multiple times for each binary in the dataset in accordance with a user-specified time or resource budget.
% After code transformations are complete for a given iteration, \TheMutator then directly swaps out the corresponding (unmodified) blocks with their transformed counterparts using the GTIRB rewriting framework. The GTIRB pretty printer~\cite{gtirb-pprinter} is then invoked to first perform the necessary bookkeeping to update memory location offsets throughout the binary, and then to reassemble the file into a new binary. The newly generated binary is then added to the original dataset and tagged with the same label (i.e., malicious or benign) as the one used during its generation. This end-to-end process repeats multiple times for each binary in the dataset in accordance with a user-specified time or resource budget.

\subsection{Code Transformations}
\label{sec:code_transformations}

\TheMutator currently supports 10 different semantics-preserving code transformations that cover the set of mutations we observed in our analysis of the popular Ember dataset (\S\ref{sec:approach}), as well as commonly used code obfuscations~\cite{llvm-obfuscator, tigress, obfuscation-survey} and transformation techniques employed by off-the-shelf optimizing compilers~\cite{dragon-book,obfuscation-wiki}. Supported transformations include junk code insertion and instruction swapping (both described in \S\ref{sec:approach}), as well as instruction substitution which replaces an instruction with a (more complex) sequence of instructions  that is semantically equivalent. To ensure that a modified code block is semantically equivalent to the original block, static analysis is performed after the code transformation is applied. This analysis tracks program reads and writes and determines whether the reads from the registers and memory locations in that basic block would still return the same values after the modification. If a violation occurs from the code transformation, it is reverted and a new transformation is attempted. Appendix \ref{sec:appendix-transforms} provides a comprehensive overview of the transformations that \TheMutator supports, as well as the logistics to carrying out each one.
%A comprehensive overview of our transformations can be found in Appendix \ref{sec:appendix-transform}.

\TheMutator's goal is to generate new versions of input binaries that differ in diverse ways from their originals while adhering to a user-specified time and/or resource budget (which dictates potential parallelism across mutation iterations).
%, in line with the transformation process that malware authors carry out to iteratively sidestep malware detectors.
The main challenge is that it is difficult to determine, a priori, how a given transformation will alter a given binary -- this depends on subtle interactions between the transformation logic and the binary instructions, which collectively dictate how many blocks are applicable for a transformation, and how many instructions will be modified, added, or deleted. Thus, during each mutation iteration, \TheMutator instead opts to randomly select multiple transformations for each mutation iteration and stochastically order them. This follows from our finding that malware authors typically employ multiple transformations together, and that binaries in the same family can differ by (largely) varying amounts (\S\ref{sec:approach}).

To further bolster variance across the transformed binaries, \TheMutator varies two parameters across the mutation iterations for each input binary. $m$  specifies the number of transformation iterations to perform on each binary, and $c$ governs the fraction of blocks to mutate in each iteration. \TheMutator maintains a running list of parameter values used for a given binary and selects subsequent values to maximize diversity, i.e., maximizing the distance from all previously used values. Note that the overarching time budget takes precedence over per-binary parameter values; to enforce this, \TheMutator round robins through the input binaries, performing one mutation iteration on each one, and circling back to fulfill the selected $m$ per binary only if time permits.

% ; note that we prioritize blocks that embed commonly used instructions such as \texttt{mov} and \texttt{add}~\cite{x86-stats}

\subsection{Optimizations for Practicality}

% \Para{Sources of inefficiency.} Binary mutation of a single executable with \TheMutator (as described thus far) can be broken down into 3 stages: (1) invoking Ddisasm to get the GTIRB representation (\emph{decompilation}), (2) carrying out semantics-preserving code transformations on instruction blocks in the GTIRB file (\emph{mutation}), and (3) converting the modified GTIRB file to assembly and generating the altered binary (\emph{reasssembly}). We profiled the runtime of each stage by passing 3K random binaries from Ember through \TheMutator.
\Para{Sources of inefficiency.} Binary mutation of a single executable with \TheMutator (as described thus far) can be broken down into 3 stages: (1) invoking Ddisasm on the binary (\emph{decompilation}), (2) carrying out semantics-preserving code transformations (\emph{mutation}), and (3) generating the output binary (\emph{reasssembly}). We profiled the runtime of each stage by passing 3K random binaries from Ember through \TheMutator.

% As shown in Figure~\ref{fig:total-pipeline}, all three stages consume substantial time: median values for the three stages across binary sizes are 0.6--33, 0.1--585, and 0.1--34 seconds, respectively. We additionally observed that per-stage delays grow as binary sizes grow. The per-stage distributions are broad, spanning upwards of 460, 961, and 44 seconds, with the precise times varying depending on the selected transformations and their interactions with each binary. Accordingly, aiming to even perform a single mutation iteration on each binary in existing small datasets (which would not fully bridge the size gap with large datasets) could take up to several thousand hours! The associated resource costs would forego the savings that practitioners reap by not purchasing existing large datasets.

As shown in Figure~\ref{fig:total-pipeline}, all three stages consume substantial time: median values for the three stages across binary sizes are 0.6--33, 0.1--585, and 0.1--34 seconds, respectively. We additionally observed that per-stage delays grow as binary sizes grow and span upwards of 460, 961, and 44 seconds. Accordingly, aiming to even perform a single mutation iteration on each binary in existing small datasets (which would not fully bridge the size gap with large datasets) could take up to several thousand hours! The associated resource costs would forego the savings that practitioners reap by not purchasing existing large datasets. Instead, \TheMutator embeds the following two optimizations to boost \TheMutator's utility for a given time budget; we evaluate the effectiveness of each one in \S\ref{sec:evaluation}, and provide more details in Appendix \ref{sec:appendix-optimizations}.

% As discussed in \S\ref{sec:approach}, fundamental to the heterogeneity in large datasets is the existence of multiple binary versions in each family and the use of combinations of code transformations to generate those versions. Thus, straightforward solutions to lower overheads (and costs) such as transforming a random subset of binaries or performing fewer transformations on each one will not suffice. Instead, \TheMutator embeds two optimizations to boost \TheMutator's utility for a given time budget: 

\indent (1) \Para{Code similarity clustering.} A clustering strategy to group binaries based on their compositions. Only a single binary per cluster is operated on, and the resulting code blocks are rapidly (but safely, from a semantics perspective) dropped into the other binaries in the same cluster. This approach circumvents costly operations for all-but-one binary per cluster, while preserving diverse interactions between code alterations and other sections in each binary.

\indent (2) \Para{Intermediate binary generation.} A technique to increase the number of diverse binaries output from each pass through the pipeline. The main difficulty is that it is difficult to (efficiently) determine, a priori, the effects that a transformation will have on a given binary's code blocks. Thus, \TheMutator opts for a dynamic approach, whereby a lightweight runtime check determines the efficacy of outputting a binary -- based on code discrepancies from the original and previously output versions -- after each transformation that is performed in a pipeline pass.

%%%%% EVALUATION %%%%%
\section{Evaluation}
\label{sec:evaluation}

%In this section, we discuss our experiments for testing the binaries produced by \TheMutator in improving malware detection performance.

% \subsection{Methodology}
\label{sec:evaluation:methodology}

To evaluate \TheMutator, we used the recent MalConv CNN-powered malware detector~\cite{malconv}. For context, MalConv's model spans 5 layers, with an embedding layer that maps bytes to vectors, and then a series of convolutional and recurrent layers. Our experiments consider 2 main datasets: (1) the high-accuracy commercial Ember dataset that includes 1.1M samples (800K after removing ill-formed binaries), and (2) the small-scale (free) Brazilian malware dataset~\cite{brazilian-malware} with 50K samples. Given the realism of Ember observed by researchers and practitioners, we use its test set, which consists of 200K benign and malicious samples, directly to reflect malware detection scenarios in the wild. For training, we consider a subset of the 600K-sample Ember training dataset, as well as the full Brazilian dataset; we train a separate MalConv model for each case. Our subsets consist of 10-30K samples, which is consistent with the dataset sizes that many malware research groups currently work with \cite{malware-datasets}. While we would have preferred to experiment with augmenting more datasets, we are constrained since many existing datasets do not contain raw binaries as mentioned in \ref{sec:background}.

%subset of the Ember training set consisting of 30K binaries (15K benign and 15K malicious) to mimic limited data available to researchers.

For each dataset, we train MalConv to convergence, routinely around 5 epochs. Training involves first collecting (converged) ``pre-trained'' weights on the original training dataset, and then running an additional training round (5 epochs) with the augmented dataset that \TheMutator generates. All training was performed on an NVIDIA PH402 with two P100s 32GB. Unless otherwise noted, \TheMutator employs combinations of all 10 of its supported transformations and generates a set of mutated binaries (split evenly across malicious and benign files); the description of each experiment specifies the number of those mutated samples considered during retraining. Accuracy is reported as the percentage of correct labels (i.e., benign or malicious) output by MalConv. We run each experiment four times and report on the distributions.

\subsection{Overall Accuracy Improvements}

\begin{figure}[t]
  \centering
  \begin{subfigure}{0.28\linewidth}

  \vspace{-1.5em}
 % This file was created with tikzplotlib v0.9.14.
\begin{tikzpicture}

\definecolor{color0}{rgb}{0.917647058823529,0.917647058823529,0.949019607843137}
\definecolor{color1}{rgb}{0.194607843137255,0.453431372549019,0.632843137254902}

\begin{axis}[
axis background/.style={fill=color0},
axis line style={white},
tick align=outside,
tick pos=both,
x grid style={white},
xlabel={Mutated Samples},
xmin=-0.5, xmax=3.5,
xtick style={color=white!15!black},
xtick={0,1,2,3},
xticklabels={3k,6k,9k,12k},
y grid style={white},
ylabel={Acc. Imp. (\%)},
ymajorgrids,
ymin=0.851975, ymax=5.37352500000001,
ytick style={color=white!15!black},
width=\columnwidth,
height=\columnwidth
]
\path [draw=white!24.7058823529412!black, fill=color1, semithick]
(axis cs:-0.4,1.614375)
--(axis cs:0.4,1.614375)
--(axis cs:0.4,2.946375)
--(axis cs:-0.4,2.946375)
--(axis cs:-0.4,1.614375)
--cycle;
\path [draw=white!24.7058823529412!black, fill=color1, semithick]
(axis cs:0.6,1.996875)
--(axis cs:1.4,1.996875)
--(axis cs:1.4,4.352375)
--(axis cs:0.6,4.352375)
--(axis cs:0.6,1.996875)
--cycle;
\path [draw=white!24.7058823529412!black, fill=color1, semithick]
(axis cs:1.6,2.011125)
--(axis cs:2.4,2.011125)
--(axis cs:2.4,3.197875)
--(axis cs:1.6,3.197875)
--(axis cs:1.6,2.011125)
--cycle;
\path [draw=white!24.7058823529412!black, fill=color1, semithick]
(axis cs:2.6,1.609)
--(axis cs:3.4,1.609)
--(axis cs:3.4,2.506875)
--(axis cs:2.6,2.506875)
--(axis cs:2.6,1.609)
--cycle;
\addplot [semithick, white!24.7058823529412!black]
table {%
0 1.614375
0 1.0575
};
\addplot [semithick, white!24.7058823529412!black]
table {%
0 2.946375
0 3.3855
};
\addplot [semithick, white!24.7058823529412!black]
table {%
-0.2 1.0575
0.2 1.0575
};
\addplot [semithick, white!24.7058823529412!black]
table {%
-0.2 3.3855
0.2 3.3855
};
\addplot [semithick, white!24.7058823529412!black]
table {%
1 1.996875
1 1.512
};
\addplot [semithick, white!24.7058823529412!black]
table {%
1 4.352375
1 5.16800000000001
};
\addplot [semithick, white!24.7058823529412!black]
table {%
0.8 1.512
1.2 1.512
};
\addplot [semithick, white!24.7058823529412!black]
table {%
0.8 5.16800000000001
1.2 5.16800000000001
};
\addplot [semithick, white!24.7058823529412!black]
table {%
2 2.011125
2 1.6245
};
\addplot [semithick, white!24.7058823529412!black]
table {%
2 3.197875
2 4.276
};
\addplot [semithick, white!24.7058823529412!black]
table {%
1.8 1.6245
2.2 1.6245
};
\addplot [semithick, white!24.7058823529412!black]
table {%
1.8 4.276
2.2 4.276
};
\addplot [semithick, white!24.7058823529412!black]
table {%
3 1.609
3 1.5625
};
\addplot [semithick, white!24.7058823529412!black]
table {%
3 2.506875
3 2.8755
};
\addplot [semithick, white!24.7058823529412!black]
table {%
2.8 1.5625
3.2 1.5625
};
\addplot [semithick, white!24.7058823529412!black]
table {%
2.8 2.8755
3.2 2.8755
};
\addplot [semithick, white!24.7058823529412!black]
table {%
-0.4 2.3
0.4 2.3
};
\addplot [semithick, white!24.7058823529412!black]
table {%
0.6 3.1195
1.4 3.1195
};
\addplot [semithick, white!24.7058823529412!black]
table {%
1.6 2.48925
2.4 2.48925
};
\addplot [semithick, white!24.7058823529412!black]
table {%
2.6 2.00425
3.4 2.00425
};
\end{axis}

\end{tikzpicture}
 \vspace{0.4em}
\caption{Accuracy on in-distribution.}
 \label{fig:main_ember}
  \end{subfigure}
  \hfill
    \begin{subfigure}{0.28\linewidth}
% This file was created with tikzplotlib v0.9.14.
\begin{tikzpicture}

\definecolor{color0}{rgb}{0.917647058823529,0.917647058823529,0.949019607843137}
\definecolor{color1}{rgb}{0.194607843137255,0.453431372549019,0.632843137254902}

\begin{axis}[
axis background/.style={fill=color0},
axis line style={white},
tick align=outside,
tick pos=both,
x grid style={white},
xlabel={Mutated Samples},
xmin=-0.5, xmax=3.5,
xtick style={color=white!15!black},
xtick={0,1,2,3},
xticklabels={3k,6k,9k,12k},
y grid style={white},
ylabel={Acc. Imp. (\%)},
ymajorgrids,
ymin=0.873205, ymax=5.420495,
ytick style={color=white!15!black},
width=\columnwidth,
height=\columnwidth
]
\path [draw=white!24.7058823529412!black, fill=color1, semithick]
(axis cs:-0.4,1.59875)
--(axis cs:0.4,1.59875)
--(axis cs:0.4,2.9905)
--(axis cs:-0.4,2.9905)
--(axis cs:-0.4,1.59875)
--cycle;
\path [draw=white!24.7058823529412!black, fill=color1, semithick]
(axis cs:0.6,2.0688)
--(axis cs:1.4,2.0688)
--(axis cs:1.4,4.373125)
--(axis cs:0.6,4.373125)
--(axis cs:0.6,2.0688)
--cycle;
\path [draw=white!24.7058823529412!black, fill=color1, semithick]
(axis cs:1.6,1.61965)
--(axis cs:2.4,1.61965)
--(axis cs:2.4,2.5381765)
--(axis cs:1.6,2.5381765)
--(axis cs:1.6,1.61965)
--cycle;
\path [draw=white!24.7058823529412!black, fill=color1, semithick]
(axis cs:2.6,2.5279)
--(axis cs:3.4,2.5279)
--(axis cs:3.4,2.841075)
--(axis cs:2.6,2.841075)
--(axis cs:2.6,2.5279)
--cycle;
\addplot [semithick, white!24.7058823529412!black]
table {%
0 1.59875
0 1.0799
};
\addplot [semithick, white!24.7058823529412!black]
table {%
0 2.9905
0 3.4432
};
\addplot [semithick, white!24.7058823529412!black]
table {%
-0.2 1.0799
0.2 1.0799
};
\addplot [semithick, white!24.7058823529412!black]
table {%
-0.2 3.4432
0.2 3.4432
};
\addplot [semithick, white!24.7058823529412!black]
table {%
1 2.0688
1 1.52910000000001
};
\addplot [semithick, white!24.7058823529412!black]
table {%
1 4.373125
1 5.2138
};
\addplot [semithick, white!24.7058823529412!black]
table {%
0.8 1.52910000000001
1.2 1.52910000000001
};
\addplot [semithick, white!24.7058823529412!black]
table {%
0.8 5.2138
1.2 5.2138
};
\addplot [semithick, white!24.7058823529412!black]
table {%
2 1.61965
2 1.5589
};
\addplot [semithick, white!24.7058823529412!black]
table {%
2 2.5381765
2 2.915371
};
\addplot [semithick, white!24.7058823529412!black]
table {%
1.8 1.5589
2.2 1.5589
};
\addplot [semithick, white!24.7058823529412!black]
table {%
1.8 2.915371
2.2 2.915371
};
\addplot [semithick, white!24.7058823529412!black]
table {%
3 2.5279
3 2.5279
};
\addplot [semithick, white!24.7058823529412!black]
table {%
3 2.841075
3 2.9265
};
\addplot [semithick, white!24.7058823529412!black]
table {%
2.8 2.5279
3.2 2.5279
};
\addplot [semithick, white!24.7058823529412!black]
table {%
2.8 2.9265
3.2 2.9265
};
\addplot [black, mark=diamond*, mark size=2.5, mark options={solid,fill=white!24.7058823529412!black}, only marks]
table {%
3 1.9369
};
\addplot [semithick, white!24.7058823529412!black]
table {%
-0.4 2.30565
0.4 2.30565
};
\addplot [semithick, white!24.7058823529412!black]
table {%
0.6 3.1708
1.4 3.1708
};
\addplot [semithick, white!24.7058823529412!black]
table {%
1.6 2.0261725
2.4 2.0261725
};
\addplot [semithick, white!24.7058823529412!black]
table {%
2.6 2.76875
3.4 2.76875
};
\end{axis}

\end{tikzpicture}
\caption{Accuracy on novel malware families only.}
 \label{fig:ember_unseen}
  \end{subfigure}
  \hfill
  \begin{subfigure}{0.42\linewidth}
 \includegraphics[width=\columnwidth]{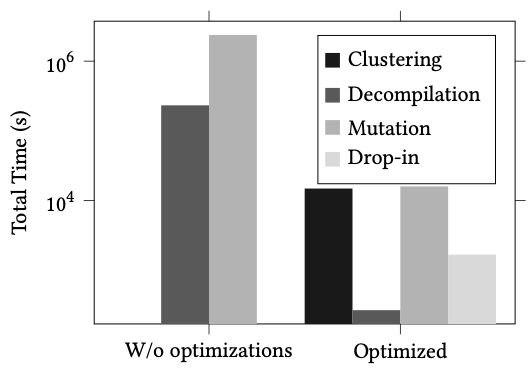}
\caption{Time in each step of \TheMutator without (left) and with (right) clustering optimization.}
 \label{fig:opt}
 \end{subfigure}
  \caption{Results of \TheMutator augmented training when testing on in-distribution data from Ember (a), novel malware families not seen in training (b). In addition the impact of our augmentations are shown in (c) and result in a two order of magnitude speedup to create the augmented training samples. }
  
 \hfill
  \caption{\TheMutator accuracy and pipeline optimization improvements}
\end{figure}

Figure~\ref{fig:main_ember} shows the accuracy improvements that \TheMutator brings to MalConv when augmenting the Ember training dataset with different numbers of mutated samples (ranging from 3-12K). Accuracy improvements range from 1--5\% atop the baseline accuracy of 61.3\% achieved when considering the unmodified Ember dataset alone. Perhaps more importantly, these results highlight that accuracy improvements typically come quickly, while operating on only a small number of binaries, e.g., adding only 3K and 6K mutated samples to the dataset delivers 3.5\% and 5\% of accuracy boosts, respectively. The reason is that \TheMutator's efficiency-centric optimizations promote rapid diversity amongst the generated samples, which in turn enable MalConv to quickly strike a desirable balance between (1) learning to detect obfuscation patterns, while (2) not overfitting to mutated samples. Results on the smaller Brazilian malware dataset \cite{brazilian-malware} were comparable: adding 2K mutated files delivered median accuracy improvements of 2\% (atop the 61\% without \TheMutator).
%Upon adding 2k mutated files, median accuracy improvements with \TheMutator were 1-2\% (atop the baseline accuracy value, without \TheMutator, of 61\%).

%the mutated versions relative to the originals (which exhibit diversity beyond simply obfuscations).

Further analysis reveals that a key driver of the overall accuracy wins delivered by \TheMutator are improvements on test samples from \emph{previously unseen} malware families, i.e., families that did not appear in the training dataset. Recall from \S\ref{sec:background} that such samples are the ones which static analysis and small-scale ML approaches typically struggle to generalize to. Figure~\ref{fig:ember_unseen} illustrates this, showing that \TheMutator's accuracy boosts on only the subset of test binaries that were not seen during training are on par with the wins on the complete test set (1--5\%). The underlying reason for these improvements is that code transformations provide a discernible pattern for MalConv to link across diverse binaries in different families. In light of these results, we provide further analysis of \TheMutator in Appendix \ref{sec:appendix-analyzing-marvolo}. 

\subsection{Pipeline Optimization Improvements}

\begin{figure}[!t]
\includegraphics[width=\textwidth]{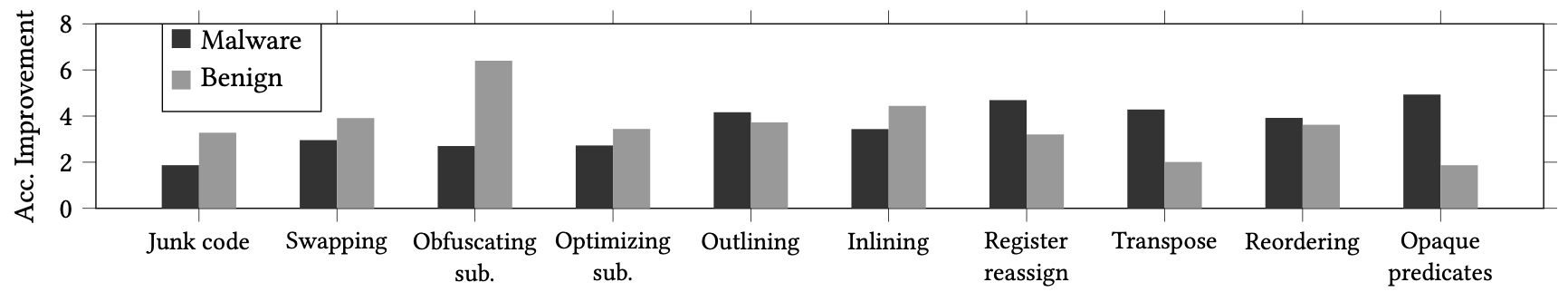}
\vspace{-1.5em}
\caption{MalConv's accuracy improvements when using a version of \TheMutator that only performs a single type of semantics-preserving code transformation during mutation. Results are for adding 1K mutated samples in each case to the Ember dataset.}
  \label{fig:param}
\end{figure}

 % Figure placed early b/c LaTeX sucks at double-wides 
%  \begin{figure*}[]
% \input{tikzplot-neurips/param-experiments}
%   \vspace{-1em}
%  \caption{MalConv's accuracy improvements when using a version of \TheMutator that only performs a single type of semantics-preserving code transformation during mutation. Results are for adding 1K mutated samples in each case to the Ember dataset.}
%   \label{fig:param}
%  \end{figure*}

 Recall from \S\ref{sec:marvolo} that \TheMutator embeds two optimizations to tackle the overheads in the mutation process revealed in our profiling results. 
%  To uncover the effects of these optimizations, 
We profiled two runs of \TheMutator's mutation pipeline, one with the two optimizations enabled, and one without them. Each pipeline was used to generate 3K mutated samples, and we note that the MalConv models trained on these mutated samples (atop the Ember dataset) delivered accuracy within 1\% of one another.  

Figure~\ref{fig:opt} shows the total time spent (i.e., to generate all 3K mutated samples) in each \TheMutator pipeline stage across these two variants. Overall, the optimized version of \TheMutator runs 79$\times$ faster to generate 3K mutated samples of similar efficacy (given the near-identical MalConv performance across the two cases noted above). Speedups are primarily from the lower decompilation and mutation/reassembly costs, which in turn are due to running only a single binary per cluster through the pipeline (85\% fewer binaries), with each run yielding a larger number of mutated samples. These drastic drops dwarf the drop-in overheads used to mix (altered) code and data blocks, and the slight (blocking) overhead of performing clustering prior to mutation; note that clustering overheads are paid once and steadily decrease in relative importance as the target number of mutated samples grows. % We used a single 192-core 384-hyperthread 4.2-TB RAM machine (Intel Xeon Processor E7-8890 v4), but did not parallelize the pipeline invocations to. 

%%%%% RELATED WORK %%%%%%

% \section{Additional Related Work}

%\subsection{Data Augmentation}

%%%% CONCLUSION %%%%%
\section{Conclusion}
\label{sec:conclusion}

\TheMutator is a data augmentation engine that boosts the efficacy of the malware datasets that practitioners commonly are restricted to by performing semantics-preserving code transformations on the constituent binaries. To the best of our knowledge, we are the first to leverage insights from a deep-dive analysis of existing malware datasets to apply meaningful data augmentation to the domain of malware detection. Key to \TheMutator's practicality are its ability to (safely) propagate labels across input and output binary samples, and its optimizations to boost the number of fruitful (i.e., diverse and representative) data samples generated within a fixed time budget. Experiments using commercial malware datasets and a recent ML-driven malware detector show that \TheMutator boosts accuracies by up to 5\%, while operating on only 15\% of the available binaries (mutation speedups of 79$\times$).

% \Para{Limitations and future work.} \TheMutator is the first viable programmatic data augmentation system for malware, but future work still remains to overcome existing limitations. First, new code alterations used by malware authors are only caught once detectors are presented with data samples or signatures that identify them. \TheMutator does not address this problem, and instead aims to maximize the utility of the data samples that practitioners have access to at any time. Second, non-trivial engineering is required for extensions to new platforms (e.g., Android).

{ \small
\bibliographystyle{IEEEtran}
\bibliography{references}
}

\appendix

\clearpage
%\section{Appendix}

\section{\TheMutator's Code Transformations}
 \label{sec:appendix-transforms}
 
 \Para{Junk code insertion.} Insert instructions into the binary that don't alter the output of the program upon being executed. These instructions may change the state of the program (e.g., register values and memory) but reverse the changes before progressing to subsequent instructions. The simplest form of this transformation that we implement is the insertion of \texttt{nop} instructions. We also generate semantic nops which consist of pushing values onto the stack, performing arithmetic and logical operations, and then popping the values off once they're completed. We augment this with additional instructions that also read and write to memory. In the following semantic nop
\begin{minted}[fontsize=\footnotesize]{gas}
                 push eax
                 inc eax
                 or  eax, 0x1c
                 add eax, dword ptr [esp - 0x34]
                 not eax 
                 pop eax 
\end{minted}
 the \texttt{eax} register is first pushed to the stack. Then arithmetic and bitwise operations are performed on \texttt{eax}. Lastly, the old value of \texttt{eax} is popped from the stack and written back into \texttt{eax}; since the value of \texttt{eax} is not written elsewhere prior to \texttt{pop eax}, the computations are effectively useless.
 \\
 \Para{Register reassignment.} Changes the names of the variables or registers. Identify a live register, \texttt{rX}, within a basic block and replace it with a new register, \texttt{rY}, that is unused within the block. The value of \texttt{rY} is first pushed onto the stack and is then written with the value stored in \texttt{rX}. After computations are performed on \texttt{rY}, it is written to \texttt{rX} and the original value of \texttt{rY} is popped and written back to \texttt{rY}.
\\
\Para{Function inlining.} Identify functions and every time they are invoked, replace the \texttt{call} instructions with the bodies of the identified functions. In our implementation, we solely focus on functions with straight-line code. Function inlining is a common compiler optimization used to reduce the overhead of invoking a function and to make basic blocks more amenable to subsequent optimizations.
\\
\Para{Function outlining.} Identify straight-line instructions within the current basic block and generate a new function with those instructions. Replace the original instructions with a \texttt{call} instruction to the newly-generated function. This is a compiler optimization for reducing code size.
\\
\Para{Obfuscating Instruction substitution.}
 Replace an instruction with a semantically equivalent sequence of new instructions. We currently support over 30 substitutions. We add simple substitutions such as changing \texttt{add rX, 1} to \texttt{sub rX, -1}. We adopt further instruction substitutions, including many implemented in LLVM Obfuscator \cite{llvm-obfuscator}. These substitutions are mostly comprised of more complex bitwise and arithmetic instructions. For instance, \TheMutator would replace the instruction \texttt{or eax,0x4711} with
\begin{minted}[fontsize=\footnotesize]{gas}
              push esi
              push edi
              mov esi, eax
              mov edi, 0x4711
              and eax, edi
              xor esi, edi
              or  eax, esi
              pop edi
              pop esi
\end{minted}
The transformation is effectively replacing $a = b | c$ with $a = (b$ \& $c) | (b \oplus c)$.
 \\
\Para{Optimizing instruction substitution.} Replace an instruction with an equivalent instruction that optimizing compilers often emit \cite{obfuscation-wiki}. While these instructions are often times not as intuitive as their more straightforward counterparts, they are faster to execute.  For instance, \texttt{mov rX, 0} is often times changed to \texttt{xor rX, rX}. Another instance is substituting arithmetic instructions, such as \texttt{add}, with \texttt{lea} instructions. Applying this transformation more broadly captures the range of programs that can be produced by different compiler toolchains and options.
\\
\Para{Code transposition.} This transformation reorders a sequence of instructions that changes the appearance of the code without altering the behavior \cite{obfuscation-survey}. \TheMutator implements code transformation by dividing a basic block into smaller slices. Then these slices are rearranged in a different order and are each appended with an unconditional \texttt{jmp} instruction to ensure that the original execution order of the initial basic block is preserved.
\\
\Para{Instruction swapping.} As another form of code transposition, we take 2 instructions and swap their positions. While this transformation does not significantly affect the readability of the code, it is used by malware authors to evade anti-virus scanners. To ensure that the transformation preserves semantics, analysis is performed to check that the swap doesn't violate any computational dependencies. We check that each of the destination registers for the instructions aren't used as a source register for other instructions. We also check that any source registers used by the two instructions aren't written to. Below we demonstrate an example; the left side shows the original program and the right side shows the modified program after the \texttt{add} and \texttt{sub} instructions had been swapped.

\begin{figure}[!h]
    \begin{minipage}{0.21\textwidth}
        \centering
        \underline{\textbf{Original}}
        \begin{minted}[fontsize=\footnotesize]{gas}
         add eax, ebx
         sub ecx, 0x7c21
         ret
    \end{minted}
    \end{minipage}
    \begin{minipage}{0.25\textwidth}
        \centering
        \underline{\textbf{Mutated}}
         \begin{minted}[fontsize=\footnotesize]{gas}
         sub ecx, 0x7c21
         add eax, ebx
         ret
        \end{minted}
    \end{minipage}
\end{figure}

On the other hand, the program
\begin{minted}[fontsize=\footnotesize]{gas}
                mov eax, 0x1af3 
                add ecx, eax
\end{minted}
is not amenable to swapping since the \texttt{add} instruction would not use the updated value in \texttt{eax} after the \texttt{mov} instruction.
\\
\Para{Opaque predicate insertion.} Opaque predicates are predicates that always evaluate to true or false and are known a priory the programmer. While opaque predicates evaluate to the same value under all inputs, they are still evaluated during runtime. To represent the instances where code and data are interleaved within a binary \cite{code-and-data}, we generate a sequence of randomly-generated bytes following the opaque predicate. An unconditional \texttt{jmp} instruction is inserted so that these generated bytes are not executed and the next instructions within the program are run. Opaque predicates are commonly inserted by code obfuscators. \cite{llvm-obfuscator}.
\\
\Para{Function reordering.} Functions are moved to different positions throughout the binary. This transformation drastically changes the appearance of the binary without adding new instructions or removing existing ones.
\\

\section{\TheMutator Optimizations}
\label{sec:appendix-optimizations}

\Para{Code similarity clustering.} To reduce the number of binaries passed through the mutation pipeline, \TheMutator employs a clustering strategy to group binaries together based on their compositions (and thus, their interactions with the pipeline). Efficiency wins come from passing only a single binary per cluster through the pipeline. Intuitively, the goal for clustering is thus to maximize cluster sizes without masking  differences between the binaries in the  dataset.

Unfortunately, the straightforward clustering strategy of grouping binaries based on byte similarity (i.e., cluster binaries whose byte-level differences are smaller than a pre-determined threshold) are ill-suited for our task. The reason is that, even malicious binaries within the same family that exhibit identical \texttt{.text} and \texttt{.data} sections may have vast byte-level differences (upwards of tens of thousands of bytes). Though massive, these differences do not alter the overall behavior of the binary, and thus should not map binaries to different clusters. Yet unearthing such insights requires passing the binary through costly decompilation, foregoing many savings.

Instead, \TheMutator leverages our finding that, within a malware family, it is not uncommon for multiple binaries to have equivalent code sections; note that these binaries commonly differ in their \texttt{.data} and \texttt{.rsrc} sections -- we discuss this below. Since \TheMutator only performs code transformations, these are the only portions of the binary that \TheMutator modifies; it is thus redundant to send binaries with identical \texttt{.text} sections through \TheMutator's pipeline. Consequently, \TheMutator operates on only a single binary per observed code section. For each generated mutated version of the binary, \TheMutator performs drop-in replacement (i.e., avoiding costly decompilation and reassembly) of the transformed code section with other binaries in the same cluster; memory location offsets are quickly updated in each affected binary. In effect, this rapidly simulates the process of passing all binaries in a cluster through the end-to-end pipeline.

Note that modifying \texttt{.data} and \texttt{.rsrc} sections in a binary may not deliver semantic equivalence. In contrast to semantics-preserving code transformations that guarantee equivalent  behavior across  program inputs, data-level modifications can alter the taken control flows in a program, resulting in  different externalized values. In light of this, \TheMutator only performs drop-in replacement for binaries in the same cluster, i.e., that have identical code sections to the one which passed through the mutation pipeline. This ensures that code-data relationships are unchanged since the same control flows would be traversed during binary execution, which in turn ensures safety in propagating labels to newly generated binaries.

\Para{Intermediate binary generation.} The goal of \TheMutator's second optimization is to maximize the useful binaries output during each pass through the mutation pipeline. Recall that, for each input binary, \TheMutator's pipeline (as described thus far) selects and performs a series of transformations to generate a single mutated binary. Thus, a simple approach to increase pipeline outputs for a given run would be to output a mutated binary after each successive transformation is performed. The issue is that the generated binaries will only differ by a single transformation pass and thus will likely fail to deliver the heterogeneity seen in large datasets; recall from  \S\ref{sec:approach} that malware authors commonly use multiple transformations to ensure substantial differences from the original malware binaries. Instead, we must ensure that the generated binaries diverge substantially from one another.

The challenge is that it is difficult to know a priori how many bytes a transformation will change in a given binary (\S\ref{sec:code_transformations}). To handle this, \TheMutator employs a lightweight runtime check after each transformation is applied to determine whether the code changes performed up until that point are comprehensive enough to warrant a new binary generation (and thus assembly). Logically, the runtime checks compare byte-level diffs between the current binary version, the original, and those output after prior transformations; if all values exceed a pre-set threshold, \TheMutator deems the current binary worthy of costly assembly (and thus, a new sample in the dataset).\footnote{While most binaries within a family have significant differences, some exhibit only minor differences between one another. Thus, \TheMutator occasionally (10\% of the time, by default) outputs binary versions even if the diff threshold has not been exceeded.} To ensure that discrepancies only pertain to behavior-affecting portions of the binary without requiring costly assembly and binary-wise diffs (which we find can consume tens of seconds), \TheMutator approximates this behavior by tracking the number of code blocks affected after each step (scaled based on the inherent intrusion level of the applied transformation~\cite{x86-stats}).

\section{Analyzing \TheMutator}
\label{sec:appendix-analyzing-marvolo}

\Para{Importance of number of binaries mutated.} Figures \ref{fig:main_ember} and \ref{fig:ember_unseen} show \TheMutator's performance as the number of added mutated binaries changes. As discussed, the benefits from \TheMutator's mutations come early from the perspective that most accuracy wins can be realized by using only a small fraction of the overall dataset as input; we observe this trend over multiple datasets of different sizes. More generally, however, \TheMutator's performance with regards to input size is collectively governed by two factors – (1) the overall dataset size, and (2) the number of input samples – that influence the relationship between the utility of malware detection insights from newly added (mutated) samples and the risk of overfitting. Intuitively, larger datasets require larger numbers of mutated samples to reap benefits because they already exhibit a sufficient amount of heterogeneity (as shown in Figure 1), and they are also far less susceptible to overfitting (as the weight of each added sample is relatively smaller).

\Para{Importance of different transformations.} To study the effect that each of \TheMutator's ten code transformations have on accuracy improvements, for each transformation (in isolation), we generated two sets of 1K mutated samples: one where all mutated samples were benign, and one where all mutated samples were malicious. Figure~\ref{fig:param} shows the accuracy improvements for MalConv running on the Ember dataset plus each of the 20 mutated datasets (one at a time). For benign files, instruction swapping, obfuscating substitutions, and function inlining yielded the largest accuracy wins, with 4\%, 6\%, and 4\% performance gains, respectively. For malicious files, register reassignment, code transposition, and opaque predicate insertion were the most fruitful with 5\%, 4\%, and 5\% performance gains, respectively. The reason is that the latter trio of transformations are more invasive (i.e., they lead to larger code alterations and resultant diffs), and are hence more often applied by malware authors to circumvent recently employed detection patterns.

Further, our results in Section 3 highlight that malware authors not only use many different kinds of code transformations, but also diverse combinations of them. Thus, \TheMutator currently opts for a general randomized selection of transformations and combinations during mutation. However, to make the most use of (limited) compute resources, a practitioner could identify which code transformations are present in the samples that they already have, and focus the augmentation process on under-represented ones.

\Para{Using \TheMutator.} Indeed \TheMutator is intended to complement existing ML-driven malware detectors and we do not propose changing hyperparameters but we recommend keeping the hyperparameter-tuning methodology the same after data augmentation. Beyond these hyperparameters, we note two additional considerations:
\begin{CompactEnumerate}
    \item \Para{Input seclection.} \TheMutator performs best when presented with inputs comprising a diverse set of binaries that differ (as the dataset allows) in family and composition, e.g., binaries with large fractions of differing code portions. Doing so aids malware detectors in identifying the underlying transformations (injected by \TheMutator) across wider-ranging contexts. Further, as noted above, \TheMutator must balance generating sufficient mutated samples to boost heterogeneity in training datasets, while avoiding overfitting to those samples. Our current implementation leverages that accuracy boosts come early (i.e., with few samples) and overfitting occurs soon after, motivating an iterative process starting with only a small number of samples.
    \item \Para{Transformation selection.} Our results in Section 3 highlight that malware authors not only use many different kinds of code transformations, but also diverse combinations of them. Thus, \TheMutator opts for a general randomized selection of transformations and combinations during mutation. However, to make the most use of (limited) compute resources, a practitioner could identify which code transformations are present in the samples that they already have, and focus the augmentation process on under-represented ones.
\end{CompactEnumerate}

\end{document}